\documentclass[aps,nofootinbib,preprintnumbers,citeautoscript,twocolumn]{revtex4}
\usepackage{amsmath,amssymb}
\usepackage{xcolor}
\usepackage{enumerate}
\usepackage[colorlinks=true,citecolor=red,linkcolor=purple,linkbordercolor=cyan]{hyperref}

\usepackage{graphicx}
\graphicspath{{/home/kapil/My_data/kapilpaper/arXiv/Herring_Flicker_Distance-Paper/arxiv_Herring_Flicker/}}

\date{\today}

\begin{document}
\title{Decoherence assisted spin squeezing generation in superposition of tripartite GHZ and W states}

\author{Kapil K. Sharma$^\ast$ and Swaroop Ganguly$^\dagger$  \\
\textit{Department of Electrical Engineering,\\
Indian Institute of Technology Bombay, Mumbai 400076, India} \\
E-mail: $^\ast$iitbkapil@gmail.com\\
E-mail: $^\dagger$swaroop.ganguly@gmail.com}
\begin{abstract}
In the present paper, we study spin squeezing under decoherence in the superposition of tripartite maximally entangled GHZ and W states. Here we use amplitude damping, phase damping and depolarisation channel. We have investigated the dynamics of spin squeezing with the interplay of superposition and decoherence parameters with different directions of the mean spin vector. We have found the mixture of GHZ and W states is robust against spin squeezing generation for amplitude damping and phase damping channels for certain directions of the mean spin vector. However, the depolarisation channel performs well for spin squeezing generation and generates permanent spin squeezing in the superposition of GHZ and W states.
\end{abstract}
\maketitle
\section{Introduction}
Spin squeezing (SS)\cite{sq1,sq2,sq3} has strong connection with entanglement\cite{e1,e2} and play the important role in quantum information theory. SS characterizes the sensitivity of quantum systems with respect to SU(2) rotations. It has its lucid applications in entanglement detection, precession enhancement in quantum meteorology, atomic clocks, Bose-Einstein Condensate states, gravitational interferometers etc.\cite{qsp}. Recently there are many studies on inequalities of spin squeezing\cite{se1,se2,se3,se4}, which are progressively used to establish mathematical criteria to detect entanglement and negative pairwise correlations in quantum systems. Most importantly spin squeezing is a good tool to study the quantum correlations in quantum ensembles, where constitutes of the system cannot be addressed individually and collective behavior of the system play the significant role, like in Bose-Einstein condensate. The framework of SS has also been used to detect the signatures of quantum chaos\cite{qc1,qc2} and identifying phase transitions in quantum systems\cite{ph1,ph2}. Primary test bench for spin squeezing is the coherent spin states (CSS)\cite{cs1,cs2}, which are states with minimal uncertainty and have lucid properties. Kitagawa-Ueda has given a mathematical criteria to study spin squeezing\cite{sq1}, which is also experimentally verified. He has been studied spin squeezing in coherent spin states (CSS) with one and two axis twisting Hamiltonians. Two-axis Hamiltonian is given as $H=\chi(J_{x}J_{y}+J_{y}J_{x})$, which is non-linear Hamiltonian and a good resource to generate spin squeezing in the system. In fact, generating spin squeezing in varieties of quantum systems by many techniques is a good area of research. The scientific community needs the generation of spin squeezing mainly for two reasons as to increase the coherence time in the system and to produce strong atom-atom interactions. There are many theoretical and experimental proposals to generate spin squeezing, these also have experimental manifestations in Bose-Einstein condensate through particle collisions \cite{gs1,gs2,gs3,gs4} and transfer of squeezing in atoms from squeezed light\cite{lg1}. It is difficult to sustain squeezing in light for a long time, so scientific community has an alternative to transfer this squeezing in atomic systems to store quantum information. This technique has the great impact to design photonic quantum memories where photos are used as an information carrier. Another technique to generate the spin squeezing, which is progressively used is through nondemolition quantum measurements. Apart from it, it is interesting to find, that decoherence also plays the role in spin squeezing generation. It is obvious that quantum systems are too evasive, such that these always interact with the environment and hence decoherence effects take place. Studies of decoherence on spin squeezing is an essential requirement. Recently the spin squeezing under various decoherence channels in CCS have been studied by X. Wang et al., they also found the phenomenon of spin squeezing sudden death (SSSD)\cite{ssd}. With the motivations from X. Wang et al. study, we have shown the spin squeezing production in GHZ and W states and find a positive impact of decoherence on these states\cite{ppi}. Spin squeezing in GHZ and W states also have been studied under particle loss and it has been found that GHZ state is very much fragile in comparison to W states\cite{dep1,dep2,dep3}.These states are important states for executing applications in quantum information and study of spin squeezing in their superposition is also important\cite{gw1,gw2}. In continuation of our study and taking the motivations from the limited studies on the effect of decoherence on spin squeezing, we proceed with the spin squeezing generation in the superposition of maximally entangled GHZ and W states in the present letter.   
\section{Spin squeezing, Decoherence channels and superposition of GHZ and W states}
In this section we present the Kitagawa\textendash Ueda (KU) spin squeezing criteria and decoherence channels under those the decoherence dynamics of superposition of GHZ and W states deal with. KU criteria is one of the most significant criteria used to characterize the spin squeezing in symmetric states under particle exchange. Here we mention that GHZ and W state are symmetric and their superposition is also symmetric, so the criteria is suitable for the present study. Mathematically KU criteria is defined as,
\begin{equation}
[(\vartriangle J_{\varphi})^{2}]_{min}\leq\frac{J}{2}. \label{e7} 
\end{equation} 
Where $J$ is the spin quantum number,  given as $J=\frac{N}{2}$ and $(\vartriangle J_{\varphi})^{2}$ is the variance of the total angular momentum along the perpendicular direction of the mean spin vector $(\vec{J}_{mean})$ in the collective system of spins. Mean spin vector play an important role to examine the spin squeezing generation in the system, it is defined as,
\begin{equation}
\vec{J}_{mean}=\left(\langle J_{x} \rangle,\langle J_{y} \rangle, \langle J_{z} \rangle\right). \label{e1} \\
\end{equation}
Where $\langle J_{i={(x,y,z)}}\rangle$ are the expectation values of angular momentum components along the $(x,y,z)$ directions respectively.
Rearranging the equation (\ref{e7}) we get
\begin{equation}
\epsilon=\frac{4[(\vartriangle J_{\varphi})^{2}]_{min}}{N}\leq 1. \label{sps}
\end{equation}
Where $N$ is the number of spins in the system and $(\epsilon)$ is the spin squeezing parameter. For unsqueezed states the criteria satisfy the condition $(\epsilon=1)$, which means there is no quantum correlation present in the state. For pure CSS uncorrelated state, the criteria carry the value as $(\epsilon=1)$. If there are certain type of quantum correlations present in the state than $(\epsilon\leq 1)$. Here we recall, KU criteria is a good tool to measure the degree of spin squeezing in superposition of GHZ and W states. The superposition of maximally entangled GHZ and W states can be written as below,
\begin{eqnarray}
|\psi\rangle=\sqrt{\alpha}|GHZ\rangle+\sqrt{(1-\alpha)}|W\rangle. \label{su}
\end{eqnarray}
with the normalization condition 
\begin{equation}
|\sqrt{\alpha}|^{2}+|(\sqrt{1-\alpha})|^{2}=1.
\end{equation}
Where,
\begin{eqnarray}
|GHZ\rangle=\frac{1}{\sqrt{2}} |000\rangle+|111\rangle \\
|W\rangle=\frac{1}{\sqrt{3}} |100\rangle+|010\rangle+|001\rangle
\end{eqnarray}
To begin with the decoherence dynamics, here we present the quantum decoherence channels, which play the important role to study the spin squeezing characteristics in Eq.\ref{su}. There are three important quantum channels widely studied in quantum information community as amplitude damping, phase damping and depolarisation channel, the corresponding  
Kraus operators of these channels are given below, 
\subsection{Amplitude damping channel}
This channel is used to  describe the energy loss in the system. The Kraus operators of amplitude damping channel are given as,
\begin{eqnarray}
E_{1}=[[1,0]^{T},[0,\sqrt{e^{\gamma t}}]^{T}]\\
E_{2}=[[0,0]^{T},[\sqrt{(1-e^{-\gamma t})},0]^{T}]
\end{eqnarray}
Where $[.]^{T}$ represents the column of the matrix.
\subsection{Phase damping channel}
This channel is the good model to represent the information loss in quantum system because of the relative phase produced in the system with system environment interaction. The channel do not involve the energy loss in the system as it is done in the case of amplitude damping channel. The corresponding Kraus operators for the channels are given as,
\begin{eqnarray}
E_{1}=[[\sqrt{e^{-\gamma t}},0]^{T},[0,\sqrt{e^{-\gamma t}}]^{T}]\\
E_{2}=[[\sqrt{(1-e^{-\gamma t})},0]^{T},[0,0]^{T}]\\
E_{3}=[[0,0]^{T},[0,\sqrt{(1-e^{-\gamma t})}]^{T}]
\end{eqnarray}
Where $\gamma t$ is the decay rate of decoherence.
\subsection{Depolarization channel}
Depolarization channel is widely studied in polarization encoding in quantum information, the corresponding Kraus operators are given as,
\begin{eqnarray}
E_{1}=[[\sqrt{e^{-\gamma t}},0]^{T},[0,\sqrt{e^{\gamma t}}]^{T}]\\
E_{2}=[[0,\sqrt{\frac{1}{3}(1-e^{-\gamma t})}]^{T},[\sqrt{\frac{1}{3}(1-e^{-\gamma t})},0]^{T}]\\
E_{3}=[[0,i\sqrt{\frac{1}{3}(1-e^{-\gamma t})}]^{T},[-i\sqrt{\frac{1}{3}(1-e^{-\gamma t})},0]^{T}]\\
E_{4}=[[\sqrt{\frac{1}{3}(1-e^{-\gamma t})},0]^{T},[0,-i\sqrt{\frac{1}{3}(1-e^{-\gamma t})}]^{T}]
\end{eqnarray}
Where $\gamma t$ is the decoherence rate and the notation $[.]^{T}$ carry the usual meaning.
\section{Results}
In this section we present the results for spin squeezing dynamics for amplitude damping, phase damping and  depolarization channels. 
\subsection{Dynamics with amplitude damping channel}
Here we present the results obtained  for spin squeezing generation with amplitude damping channel. We recall that the direction of mean spin vector play an important role to examine the spin squeezing generation, its direction can be represented on the surface of unit Bloch sphere with the combinations of angles $(\theta,\phi)$. We have plotted the spin squeezing parameter $(\epsilon)$ given in Eq.\ref{sps} with the mean spin vector directions $\{\theta\in[0,90],\phi\in [0,180]\}$. In Eq.\ref{su}, $(\alpha=0)$ corresponds to W states and $(\alpha=1)$ corresponds to GHZ states. Here we present the results in a table with the parameters $(\theta,\phi)$ for which there is no SS generation found in superposition of GHZ and W states with $(\forall \alpha)$.
\begin{center}
\begin{tabular}{|c|c|c|c|c|c|c|c|c|c|c|}
\hline 
$\theta$ & $0$ & $0$ & $0$ & $0$ & $0$ & $0$ & $0$ & $30$ & $90$ & $90$\tabularnewline
\hline 
\hline 
$\phi$ & $0$ & $30$ & $60$ & $90$ & $120$ & $150$ & $180$ & $0$ & $0$ & $180$\tabularnewline
\hline 
\end{tabular}
\end{center}
Looking into the table we have seen as the mean spin vector is along the z direction ie. $(\theta=0)$ than with $(\forall \phi,\forall \alpha)$, no spin squeezing generation takes place. It is because, in Eq.\ref{sps} the term $(\vartriangle J_{\varphi})^{2}]_{min})=0$ , so it represents that, the minimum variance along the perpendicular direction of mean spin vector $(\vec{J}_{mean})$ falling in XY plane is zero and decoherence do not induce the fluctuations in the system. So as long as the mean spin vector is along the z axis, the superposition of GHZ and W states is robust to support spin squeezing generation. Similar conclusions can be drawn from another combinations of parameter values $(\theta,\phi)$ presented in the table. 

\begin{figure*}
\centering
     \includegraphics[scale=0.6]{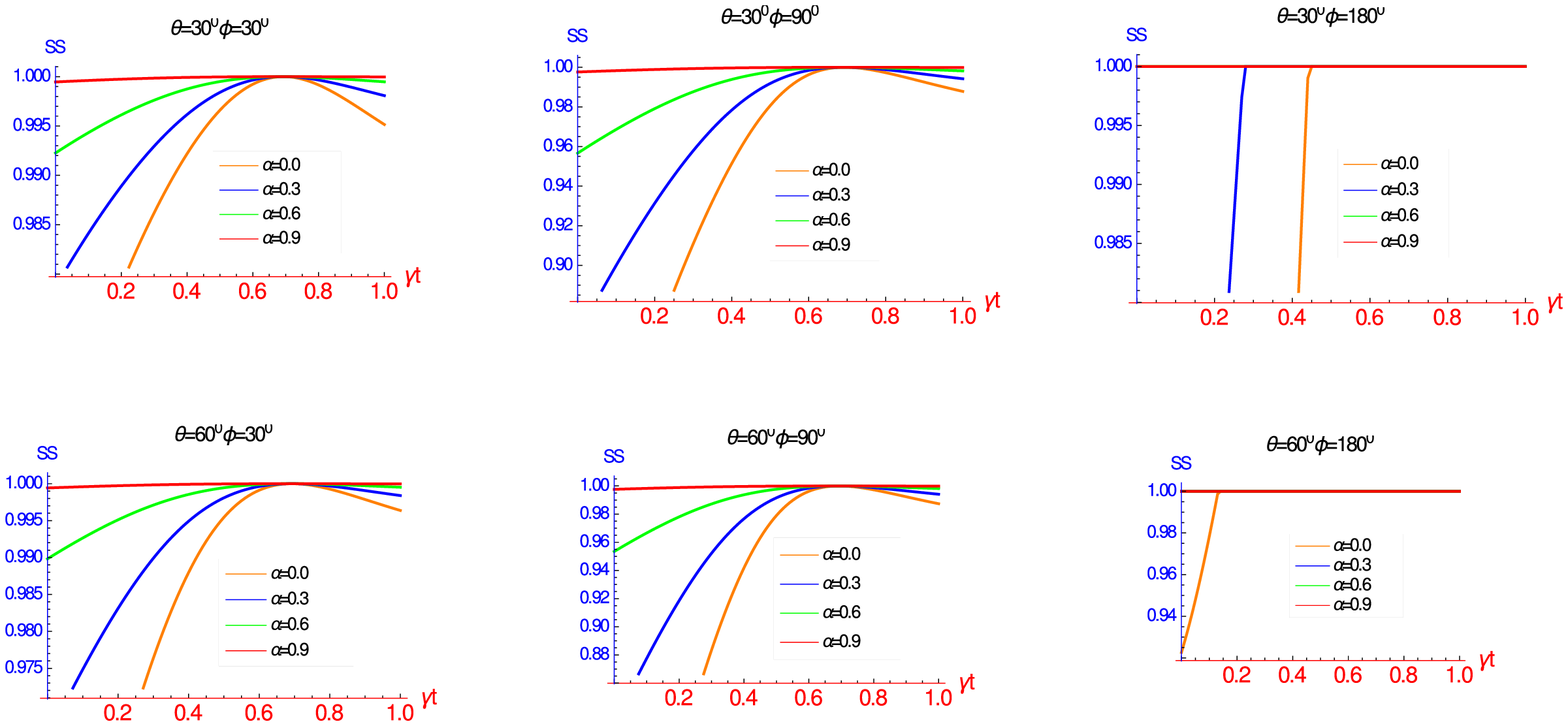}
       \caption{Spin squeezing plots for amplitude damping channel}\label{ampf}
\end{figure*}

Further results for different values of parameters $(\theta,\phi)$, which support spin squeezing generation with the effect of decoherence $(\gamma t)$ have been shown in Fig.\ref{ampf}. One important result we have found that, for $(\alpha=0.9)$, the parameter $( \epsilon=1)$ and the state $|\psi\rangle$ remain unsqueezed with this value.
\subsection{Dynamics with phase damping channel}
\begin{figure*}
\centering
     \includegraphics[scale=0.6]{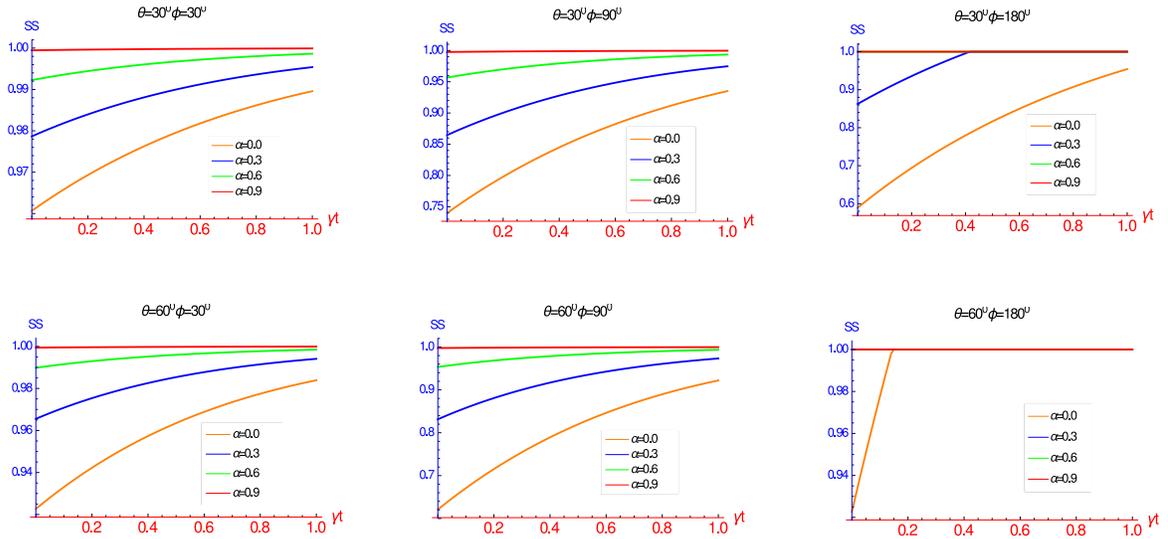}
       \caption{Spin squeezing plots for phase damping channel}\label{phasef}
\end{figure*}

Here we present the results for spin squeezing generation with phase damping channel under the effect of decoherence. First, we present those values of parameters $(\theta,\phi)$ for which the spin squeezing generation does not take place in the superposition state given in Eq.\ref{su}. The table is given below,
\begin{center}
\begin{tabular}{|c|c|c|c|c|c|c|c|c|c|c|}
\hline 
$\theta$ & $0$ & $0$ & $0$ & $0$ & $0$ & $0$ & $0$ & $30$ & $90$ & $90$\tabularnewline
\hline 
\hline 
$\phi$ & $0$ & $30$ & $60$ & $90$ & $120$ & $150$ & $180$ & $0$ & $0$ & $180$\tabularnewline
\hline 
\end{tabular} 
\end{center}
Looking into the table, here we find similar results as obtained for amplitude damping channel. So for the values given in the table, both the amplitude and phase damping channels affect the system in the same way and do not generate spin squeezing. However, the spin squeezing behavior except the values given in the table is different for phase damping channel, which is shown in Fig.\ref{phasef}. Looking in these results, we find there are nice signatures for spin squeezing generation, but the state $|\psi\rangle$ remain unsqueezed with the parameter value $(\alpha=0.9)$, which is the same result obtained for amplitude damping channel.
\subsection{Dynamics with depolarization channel}
\begin{figure*}
\centering
     \includegraphics[scale=0.6]{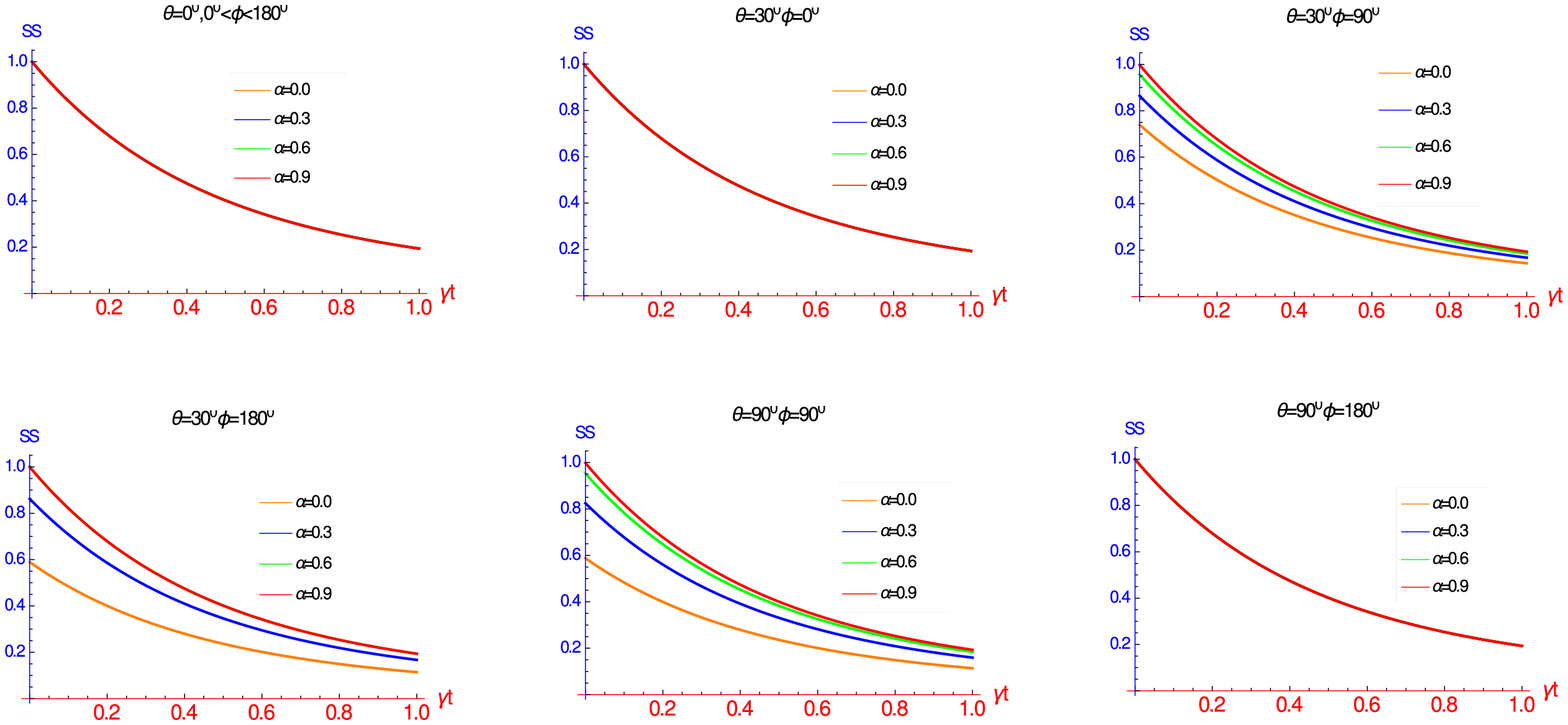}
       \caption{Spin squeezing plots for depolarization channel}\label{depolf}
\end{figure*}

In continuation with the study, here we present the results obtained for depolarization channel. This channel has a totally different impact on spin squeezing generation in comparison to amplitude damping and phase damping channel. It generates the spin squeezing in the system even if the mean spin vector is along the z-axis with $(\theta=0, 0\leq \phi\leq 180)$ for $(\forall \alpha)$. It is an opposite case to amplitude damping and phase damping channel. Looking into the graphical results shown in Fig.\ref{depolf}, we find as the values of the parameters $(\theta,\phi)$ vary, there are variations in initial amplitude of spin squeezing, but as the decoherence parameter $\gamma t$ advances, the spin squeezing decay slowly, but never become zero. So the channel generates permanent spin squeezing in the state $|\psi\rangle$. It is valuable to mention that, the decoherence channel contributes well for spin squeezing generation in the superposition of GHZ and W states.

\section{Conclusion}
In the present article, we have discussed the effect of decoherence on the superposition of GHZ and W states under amplitude damping, phase damping, and depolarization channels. For these channels, we have investigated the lucid signatures of spin squeezing generation. We also have investigated the sensitive directions of the mean spin vector for which the spin squeezing generation do not take place. The superposition parameter $(\alpha=0.9)$ has been found sensitive, which do not support spin squeezing generation in both amplitude and phase damping channel. Amplitude damping and phase damping channels have an almost same impact on spin squeezing, while depolarization channel has a totally different paradigm. For depolarization channel, as the decoherence rate increases, the degree of squeezing slowly decay but never be zero. So we have found the depolarization channel is a good resource for generating permanent spin squeezing under decoherence in the superposition of GHZ and W states. The present study can also be upgraded to study the spin squeezing generation under particle loss along with decoherence in the superposition of GHZ and W states. So we mention the results obtained in the paper and discussion may be useful for quantum information processing community. 
\section{Acknowledgement}
The authors acknowledge support from the Ministry of Electronics \& Information Technology, Government of India, through the Centre of Excellence in Nanoelectronics, IIT Bombay.


\begin{thebibliography}{99}
\bibitem{sq1}
M. Kitagawa and M. Ueda, Squeezed spin states, Phys. Rev. A \textbf{47}, 5138 (1993).
\bibitem{sq2}
D. J. Wineland, J. J. Bollinger, W. M. Itano, F. L. Moore, and D. J. Heinzen, Spin squeezing and reduced quantum noise in spectroscopy, Phys. Rev. A \textbf{46}, R6797 (1992).
\bibitem{sq3}
D. J. Wineland, J. J. Bollinger, W. M. Itano, and D. J. Heinzen, Squeezed atomic states and projection noise in spectroscopy, Phys. Rev. A \textbf{50}, 67 (1994).

\bibitem{e1}
A. Einstein, B. Podolsky, N. Rosen: Can quantum-mechanical description of physical reality be considered complete? Phys. Rev.\textbf{47}, 777 (1935).

\bibitem{e2}
M.A Nielsen.,  I.L Chuang.: Quantum Computation and Quantum Information. Cambridge University Press, Cambridge (2000).

\bibitem{qsp}
J. Maa, X. Wanga, C.P. Suna, F. Nori, Quantum spin squeezing, Phy. Reports, \textbf{509}, 89 (2011).

\bibitem{se1}
J. K.Korbicz, J. I. Cirac and M. Lewenstein, Spin squeezing inequalities and entanglement of N qubit states Phys. Rev. Lett. \textbf{95} 120502 (2005).

\bibitem{se2}
J. K. Korbicz, O. Gühne, M. Lewenstein, H. Häffner, C. F. Roos and R. Blatt, Generalized spin-squeezing inequalities in N-qubit systems: theory and experiment Phys. Rev. A \textbf{74} 052319  (2006).

\bibitem{se3}
G. Toth, . Knapp, O. Gühne and H. J. Briegel, Spin squeezing and entanglement Phys. Rev. A \textbf{79} 042334 (2009).

\bibitem{se4}
I. Saideh, S. Felicetti, T. Coudreau, P. Milman and A. Keller, Generalized spin-squeezing inequalities for particle number with quantum fluctuations Phys. Rev. A \textbf{94} 032312 (2016).

\bibitem{qc1}
E. J. Heller, Bound-state eigenfunctions of classically chaotic Hamiltonian systems: Scars of periodic orbits, Phys. Rev. Lett. \textbf{53}, 1515 (1984).

\bibitem{qc2}
R. Schack, G. M. D'Ariano, and C. M. Caves, Hypersensitivity to perturbation in the quantum kicked top, Phys. Rev. E \textbf{50}, 972 (1994).

\bibitem{ph1}
J. Vidal, Phys. Rev. A 73, 062318 (2006).

\bibitem{ph2}
J. Vidal, G. Palacios, and R. Mosseri, Entanglement in a second-order quantum phase transition, Phys. Rev. A \textbf{69}, 022107 (2004).


\bibitem{cs1}
S. T. Ali, J. P. Antoine, J. P. Gazeau and U. A. Mueller, Coherent states and their generalizations: a mathematical overview Rev. Math.
Phys. \textbf{7} 1013 (1995). 

\bibitem{cs2}
R. Holtz  and J. Hanus, On coherent spin states J. Phys. A: Mathematical Nuclear and General \textbf{7} \textbf{4} (1974).

\bibitem{gs1}
C. Orzel, A. Tuchman, M. Fenselau, M. Yasuda, and M. Kasevich, Squeezed states in a Bose-Einstein condensate, Science \textbf{291}, 2386 (2001).

\bibitem{gs2}
J. Esteve, C. Gross, A. Weller, S. Giovanazzi, and M. K. Oberthaler, Squeezing and entanglement in a Bose-Einstein condensate, Nature \textbf{455}, 1216 (2008).

\bibitem{gs3}
C. Gross, T. Zibold, E. Nicklas, J. Esteve, and M. K. Oberthaler, Nonlinear atom interferometer surpasses classical precision limit, Nature \textbf{464}, 1165 (2010).

\bibitem{gs4}
M. F. Riedel, P. Boehi, Y. Li, T. W. Haensch, A. Sinatra, and P. Treutlein, Atom-chip-based generation of entanglement for quantum metrology, Nature \textbf{464}, 1170 (2010).

\bibitem{lg1}
J. Hald, J. S$\phi$rensen, C. Schori, and E. Polzik, Spin squeezed atoms: A macroscopic entangled ensemble created by light,
Phys. Rev. Lett. \textbf{83}, 1319 (1999).

\bibitem{ssd}
X. Wang, A. Miranowicz, Y. Liu, C. P. Sun and F. Nori, Sudden vanishing of spin squeezing under decoherence Phys. Rev. A \textbf{81} 022106 (2010).

\bibitem{ppi}
Kapil K. Sharma and Swaroop Ganguly, Positive impact of decoherence on spin squeezing in GHZ and W states, J. Phys. Commun., \textbf{2}  015012 (2018).

\bibitem{dep1}
J. Stockton, J. Geremia, A. Doherty, and H. Mabuchi, Characterizing the entanglement of symmetric many-particle spin-1/2 systems, Phys. Rev. A \textbf{67}, 022112 (2003).

\bibitem{dep2}
A. Micheli, D. Jaksch, J. Cirac, and P. Zoller, Many-particle entanglement in two-component Bose-Einstein condensates,
Phys. Rev. A \textbf{67}, 013607 (2003).

\bibitem{dep3}
Y. Li, Y. Castin, and A. Sinatra, Optimum spin squeezing in Bose-Einstein condensates with particle losses, Phys. Rev.
Lett. \textbf{100}, 210401 (2008).

\bibitem{gw1}
X. J.  Yi, J. M. Wang, Spin Squeezing of Superposition of Multi-Qubit GHZ State and W State, Int. J. Theor. Phys \textbf{50} 2520 
 (2011). 
 
\bibitem{gw2} 
X. J. Yi, G. Q. Huang, J. M. Wang, Spin Squeezing in Superposition of GHZ State and W States, Int. J. Theor. Phys. \textbf{51} 2960 (2012).
\end{thebibliography}
\end{document}